\documentstyle[11pt, aaspp4,epsf,rotate]{article}

\def\ugr{\, \lower4pt \hbox{$\buildrel > \over \sim$} \, }
\def\ukl{\, \lower4pt \hbox{$\buildrel < \over \sim$} \, }

\lefthead{B\"ottcher \& Liang}
\righthead{Comptonization signatures in the rapid aperiodic variability
of GBHCs}

\begin{document}

\title{A New Model for the Hard Time Lags in Black Hole X-Ray Binaries}
\author{M. B\"ottcher \& E. P. Liang\altaffilmark{1}}
\altaffiltext{1}{Rice University, Space Physics and Astronomy Department, MS 108\\
6100 S. Main Street, Houston, TX 77005 -- 1892, USA}

\bigskip
\centerline{\it To appear in Astrophysical Journal Letters}
\bigskip

\begin{abstract}
The time-dependent Comptonized output of a cool soft X-ray source 
drifting inward through an inhomogeneous hot inner disk or corona 
is numerically simulated. We propose that this scenario can explain 
from first principles the observed trends in the hard time lags 
and power spectra of the rapid aperiodic variability of the X-ray 
emission of Galactic black-hole candidates.
\end{abstract}


\section{Introduction}

The rapid aperiodic variability of the X-ray emission of 
Galactic black-hole candidates (GBHCs) and low-mass X-ray 
binaries (for a review see \cite{vdk95}) is attracting an
increasing amount of attention since spectral information
alone is generally not sufficient to distinguish between 
different models for the high-energy emission of X-ray 
binaries. Its power spectrum density (PSD), autocorrelation 
function, time lags between different energy bands, coherence 
etc. provide important diagnostics of important parameters 
such as the extent of the emitting region, the dominant 
microscopic timescales, the dominant emission mechanism 
and the geometry of the source (\cite{kht97}, \cite{hkt97}, 
\cite{bl98}).

Early measurements of the Fourier-frequency dependence of 
time lags between the signals in different X-ray energy channels 
from GBHCs (\cite{mkk88}, \markcite{mik93}1993) have been 
interpreted as evidence against Comptonization of soft photons 
in a hot, uniform plasma as the mechanism for the production 
of the rapidly variable hard X-rays. Recently, Kazanas et al. 
(\markcite{kht97}1997) and Hua et al. (\markcite{hkt97}1997) 
have pointed out that Comptonization of internal soft photons 
in an inhomogeneous hot corona can reproduce the observed 
PSD slopes and hard time lags of Cyg~X-1 and other Galactic 
black-hole candidates in the low-hard state. B\"ottcher \& 
Liang (\markcite{bl98}1998) have expanded this idea further 
and discussed observable signatures to distinguish between 
different source geometries. 

However, B\"ottcher \& Liang (\markcite{bl98}1998) also pointed 
out that any scenario in which the observed hard time lags are 
purely due to static Comptonization, requires that the radial 
extent of the hot corona exceeds $\sim 10^4$ Schwarzschild radii 
of a solar-mass black hole ($R \gtrsim 10^{10}$~cm). This is 
incompatible with current models of accretion flows onto Galactic 
black holes (for a recent review see \cite{liang98}) even from
simple energy arguments. Moreover, according to Vaughan \& Nowak 
(\markcite{vn97}1997) a pure Comptonization scenario might have 
problems reproducing the observed coherence functions due to the 
stochastic nature of the Comptonization process, and Hua et al. 
(\markcite{hkt97}1997) mention that the coherence observed in 
Cyg~X-1 can only be reproduced by Comptonization in an inhomogeneous 
hot corona if the physical conditions of the Comptonizing region 
remain constant over long time scales (hours). This suggests
that additional variability of the physical parameters of the 
Comptonizing region may be involved in the rapid aperiodic 
variability of GBHCs.

A first model along these lines was developed by Poutanen \&
Fabian (\markcite{pf98}1998) who suggested that the Comptonization 
occurs in active regions in an optically thin corona above a cold 
accretion disk (\cite{hm91}, \markcite{hm93}1993, \cite{stern95}). 
The active regions were assumed to be heated, e. g., due to magnetic 
flares (\cite{nm97}), causing the coronal temperature to rise 
exponentially with time before it drops  off sharply. With these 
temperature profiles, the observed X-ray spectra and hard time 
lags of Cyg~X-1 could be reproduced. However, the flare 
time profile, in particular the instantaneous temperature 
drop at the end, required to yield the observed hard time lags,
seems rather artificial. There also seems to be a problem with
the coherence function in such a scenario since it would require
that the entire corona flares uniformly and homogeneously.

In this Letter, we propose an alternative and, we believe, more
natural model in which a blob of cool, dense matter spirals inward 
through an inner hot corona until it disappears at the event horizon 
of the black hole. The inner portions of the accretion flow are 
assumed to form a quasi-spherical, inhomogeneous, hot, optically 
thin corona, which could either be a Shapiro-Lightman-Eardley 
(\markcite{sle76}1976) type accretion disk or an advection 
dominated or transonic accretion flow (\cite{ny94}, \cite{cal95}, 
\cite{esin97}, \cite{luo98}). This picture is
partially motivated by the 2-phase disk scenario recently 
proposed by Krolik (\markcite{krolik98}1998). As the cool, 
dense blob drifts inward, the soft X-ray radiation it emits 
is Comptonized in an increasingly hot and dense corona, which 
results in the Comptonized X-ray spectrum becoming harder 
with time, causing the observed hard time lags. The key idea
here is that {\it the observed cutoff period of seconds} in the 
hard-time-lags versus Fourier-period plots (e. g., \cite{mkk88},
\cite{cui97a}) {\it is associated with the radial drift time scale}
of the blob.

Details of the numerical scheme adopted to simulate the
above scenario as well as an extensive parameter study
will be published elsewhere (B\"ottcher \& Liang 1999, 
in preparation). The Monte-Carlo code used here has been
described in B\"ottcher \& Liang (\markcite{\bl98}1998).
In Section 2, we describe the model and outline the method 
of our numerical simulations. Resulting power spectra and 
time lags are presented and discussed in the light of 
observational results in Section 3. Section 4 contains a 
short summary and our conclusions.

\section{The model}

The basic model presented here consists of a soft X-ray source 
(the ``blob''), emitting a thermal blackbody spectrum of 
temperature $T_{BB}$ with $0.1$~{\rm keV}~$\lesssim \, k T_{BB} 
\, \lesssim 1$~keV, which is drifting inward with a constant 
radial drift velocity $\beta_r \, c$. $T_{BB}$ is allowed to 
have a power-law dependence $T_{BB} \propto r^{-p_{BB}}$ on 
the distance $r$ from the center of the black hole. When the 
soft photon source reaches the radius $r_{in}$ of the last stable 
orbit of the accretion disk, it rapidly falls onto the event horizon 
of the black hole and disappears. In the inner $\sim 100 \, R_S$ 
around the black hole, we assume the existence of a quasi-spherical, 
hot, inhomogeneous corona. For the simulations presented in this 
Letter, we fix the radial dependence of the density $n_c$ and 
temperature $T_c$ in this central corona to values 
representative of an advection dominated accretion 
flow (\cite{ny94}, \cite{cal95}), i. e., a temperature 
profile $T_c \propto r^{-1}$ and a density profile 
$n_c \propto r^{-3/2}$ are assumed.

To calculate the observable spectra and temporal variability 
resulting from this model we perform time-dependent Monte-Carlo 
simulations of Comptonization in an inhomogeneous medium. The 
Comptonization Code we use is based on the code developed by 
Canfield et al. (\markcite{chp87}1987, see also \cite{liang93} 
and \cite{bl98}). The radial profile of the corona is approximated 
by dividing the corona into $\sim 50$ radial zones of approximately 
equal Thomson depth increments. Within each zone the density and 
temperature have a constant value. The inner boundary of the corona 
is treated as a purely absorbing boundary (the black hole horizon). 
We are simulating a single blob drifting inward through 
this corona. The photons escaping the corona at its 
outer boundary $r_{out} \sim 100 \, R_S$ are sampled
in small time bins ($\Delta t \lesssim 0.01$~s) and
in 5 energy bins. The resulting light curves in all
energy bins are then Fourier transformed, and the
corresponding power spectra and hard time lags are
calculated.

A rough analytical estimate of the mean energy of escaping 
photons as a function of escape time can be found based on 
the mean number of scatterings, $\overline N_{sc} (r_{em}) 
\approx \max\lbrace \tau_T, \tau_T^2 \rbrace$, where $\tau_T$ 
is the radial Thomson depth of the corona from the point of 
emission of a soft photon at radius $r_{em}$, to the outer 
boundary. Due to the steep density gradient, most of the 
Comptonization will take place around $r_{em}$. Thus, the 
escaping photon energy is $E_{esc} \approx E_{em} \, 
e^{4 \Theta_c (r_{em}) \overline N_{sc}}$, where $E_{em}
\approx 2.8 \cdot k T_{BB} (r_{em})$ is the average
energy of a photon emitted by the blob, and $\Theta_c
(r_{em}) = k T_c (r_{em}) / (m_e c^2)$. The time $t_{esc}$
at which a photon emitted at $r_{em}$ escapes from the
corona, may be estimated by $t_{esc} \approx t_{em} + 
(r_{out} - r_{em}) \, \sqrt{\overline N_{sc} + 1} \> / c$. 
The general trend resulting from this estimate is consistent 
with the observed approximate dependence $\Delta t_{kl} 
\propto \ln (E_k / E_l)$ (\cite{mkk88}, \cite{nowak98}) of 
the hard time lags $\Delta t_{kl}$ on the mean photon energies 
$E_{k,l}$ in the energy channels $k$ and $l$, respectively.

\section{Numerical results and discussion}

We have performed a series of simulations with coronal parameters
fixed to the ADAF self-similar profiles, but different assumptions 
on the blob temperature and its evolution. We find that the best 
agreement with observed features in the power spectrum and hard 
time lags of X-ray binaries is achieved if the blob heats according 
to $T_{BB} \propto r^{-1/4}$. The power spectra of the X-ray signals
in the different energy bands and hard time lags between adjacent
energy bins resulting from a simulation with $kT_{BB} = 0.2$~keV 
at the last stable orbit of the accretion flow around a Schwarzschild
black hole are shown in Figs.~1 and 2. 

The power spectra are characterized by a flat plateau below 
some break frequency and turn into a power law with index
$\approx 2$ at higher frequencies. The break frequency increases with
increasing photon energy, from $\ll 0.1$~Hz at $\lesssim 1$~keV
to $\sim 2$~Hz in the 25 -- 100~keV band. This is in excellent
agreement with the observed power spectrum of Cyg~X-1 (\cite{cui97a},
\markcite{cui97b}1997b, \cite{nowak98}) which turns into a power law 
of index $\approx 2$ above a break frequency of a few Hz. No significant 
dependence of the high-frequency slope on the photon energy was 
observed by Nowak (\markcite{nowak98}1998) for Cyg~X-1 and by Motch 
et al. (\markcite{mrp83}1983) and Maejima et al. (\markcite{mmm84}1984)
for GX~339-4. A flattening of the PSD with increasing photon energy, 
which is occasionally detected (\cite{bps74}, \cite{mk89}) could be 
caused by the increase of the break frequency in the PSD, evident in 
Fig. 1. We point out that the simplistic scenario simulated here is
only relevant to the high-frequency portion of the power spectrum, 
corresponding to variations on short time scales. On longer time 
scales (lower frequencies) the power spectrum in reality is dominated 
by the time series of multiple blobs which might correspond to accretion 
instabilities in the outer accretion disk. The observed broadband PSD is
simply the product of the single-blob PSD calculated here and the power
spectrum of the time series of blob recurrences.

The hard time lags between adjacent energy bands are generally 
linearly proportional to the Fourier period, turning into a flat 
curve beyond a break period, which decreases with increasing photon 
energy. For the highest energy channels, the time lags depend 
on the Fourier period to a power less than 1. The maximum time 
lags between medium-energy X-ray bands (in the energy range 
accessible to the PCA onboard RXTE) are of the order $\sim 0.1$~s. 
This is in very good agreement with the hard time lags observed
in Cyg~X-1 (\cite{mkk88}, \cite{cui97b}, \cite{nowak98}). Even
the flattening of the time lag curves towards $\Delta t \propto
t^{-0.8}$ as observed by Crary et al. (\markcite{crary97}1997) 
at high photon energies is well reproduced by our simulations.

The photon spectrum produced by this scenario consists of a 
thermal hump peaking around $\sim 0.2$ -- $0.3$~keV and a power 
law of photon index $\alpha \sim 1.5$, typical of GBHC low-hard
state spectra. Since we may assume that the global parameters of 
the corona and of the evolution and motion of the blob do not 
dramatically change on short time scales, the coherence function 
(\cite{vn97}) is close to 1 for most frequencies, except at very 
high frequencies, where photon counting noise will destroy the 
coherence, and at very low frequencies, where the stochastic 
sequence of blob excitation events will produce an incoherent 
variability pattern. Such a high degree of coherence over a broad 
Fourier frequency range has indeed been observed for Cyg~X-1 
(\cite{cui97b}, \cite{nowak98}). 

When we increase the heating rate of the inward-moving blob 
we find basically the same trend as described above for the 
power spectra, but the hard time lags turn into much steeper 
curves, approaching $\Delta t \propto P_F^2$ (where $P_F$ is 
the Fourier period) for a radial blob temperature dependence 
$T_{BB} \propto r^{-3/4}$ (\cite{ss73}). The temperature of 
the blob at the last stable orbit is tightly constrained by 
the observed soft excess in GBHC X-ray spectra, indicating a 
typical blackbody temperature around $kT_{BB} \sim 0.2$~keV. 
If $T_{BB} (r_{in})$ differs significantly from this value,
the resulting X-ray spectrum becomes inconsistent with the 
observed spectra of Cyg~X-1 and other GBHCs in the low --- 
hard state. 

The maximum time lags and the turnover frequencies in
the PSD and the time lag curves are determined by the
drift time scale of the blob through the innermost, hottest
part of the corona. All observables resulting from our model 
are insensitive to the value of the outer radius of the
ADAF-like corona because the outer regions of the corona
are inefficient in terms of Comptonization. With increasing 
drift velocity the maximum time lags become smaller, the 
turnover in the PSDs is shifted towards higher frequencies,
and the turnover in the time lag curves toward a constant
value occurs at shorter Fourier periods. If the radial
drift time scale of the blob approaches the photon escape
time scale for photons emitted near the black hole horizon,
the resulting light curves become almost symmetric, and the
hard time lags become small or even negative. Thus the choice
of a radial drift velocity $\beta_r \lesssim 1$~\%, much 
smaller than the free-fall velocity, is essential for our
model to reproduce the observed trends in PSDs and hard
time lags.

A detailed parameter study, combined with detailed fitting
of the observed X-ray spectra, PSDs, time lags and coherence
functions of Cyg~X-1 and GX~339-4 will be presented in a 
forthcoming paper.

In this Letter, we have only presented results from a single
blob. However, in reality the observed X-ray signal will
consist of the contributions from multiple blobs formed
in an irregular sequence. Our results may be regarded as
a transfer function $tr_k(t)$ describing the X-ray response 
of the corona in photon energy channel $k$ to a single blob 
moving inward. The effects of the convolution of a time sequence 
of elementary events with such a response function was briefly 
mentioned in B\"ottcher and Liang (\markcite{bl98}1998) 
and demonstrated in detail in Nowak et al. (\markcite{nowak99}1999). 
Let $b(t)$ be the number of identical blobs entering the corona 
per unit time. Then the observed signal in energy channel $k$ will 
be $f_k (t) = (b \ast tr_k)(t)$, where $\ast$ denotes the convolution. 
The Fourier transform of $f_k (t)$ is $F_k (\nu_F) = B(\nu_F) \,
Tr_k (\nu_F)$, where capital letters denote the Fourier transforms
of the respective functions. Since $b(t)$ is independent of the
photon energy, the resulting time lags are exactly given by the
single-blob time lags as plotted in Fig. 2. The resulting power 
spectrum is $PSD_k (\nu_F) = \vert B(\nu_F) \vert^2 \, \vert 
Tr_k (\nu_F) \vert^2$. The observed PSDs of GBHCs are 
typically given by a flat shape below a break frequency
$\nu_1 \sim 0.1$~Hz, a power-law $PSD \propto \nu_F^{-1}$
for $\nu_1 \lesssim \nu_F \lesssim \nu_2 \sim 2$~Hz, and a
steeper power-law $\propto \nu^{-\alpha}$, often consistent
with $\alpha \sim 2$ for $\nu_F \gtrsim \nu_2$ (e. g., 
\cite{cui97c}, \cite{rut98}, \cite{nowak98}, \cite{grove98}).
Thus, comparing to Fig. 1, we can deduce that the intrinsic
power spectrum of the blob formation events, $B(\nu_F)$,
must be roughly flat below $\nu_1$, a power-law $\propto
\nu_F^{-1}$ for $\nu_1 \lesssim \nu_F \lesssim \nu_2$,
and again flat for $\nu_F \gtrsim \nu_2$. This 
indicates that blob formation events happen on typical
repetition time scales $\sim 1$~s~$ \lesssim \Delta t_{rep}
\lesssim 10$~s with an intrinsic power spectrum $\vert B(\nu_F) 
\vert^2 \propto \nu_F^{-1}$. 

Another important point is the question of stability of the
cool blobs since our simulations assume that the blobs are
heating up slowly, but are maintaining their integrity. In
the context of the inhomogeneous accretion equilibrium 
solution of Krolik (\markcite{krolik98}1998), this is
equivalent to the condition that thermal conduction from
the hot to the cold phase is negligible, which yields a 
constraint on the minimum extent $R_b$ of the cool blobs, 
namely $R_b \gtrsim 1.73 \, h \, \tau_c^{-1} \, \Theta_c^{3/2}$, 
where $h$ is the scale height of the hot phase, $\tau_c$ is 
its half Thomson depth, and $\Theta_c = k_B T_c / (m_e c^2)$ 
is its dimensionless temperature. However, in the case
considered here, we only need to ensure the weaker condition 
that the time scale of conductive heating and thus cloud
evaporation be longer than the drift time scale of the clouds
through the innermost, hot region of the corona, which is 
$\sim 1$~s. A detailed time-dependent analysis of the stability
of spherical clouds to evaporation, including the effects of
viscous stresses, but neglecting radiative heating and cooling, 
has been presented by Draine \& Giuliani (\markcite{dg84}1984). 
If viscous stresses are unimportant, the fractional mass loss 
rate due to thermal conduction is

\begin{equation}
{\dot M_B \over M_B} = {12 \over 25} {\kappa_{class} \over
k_B \, R_B^2 \, n_{B}} \approx 8 \cdot 10^{-3} \, \left(
{k_B T_c \over 100 \, {\rm keV}} \right)^{5/2} \, R_7^{-2} 
\, n_{20}^{-1} \; {\rm s}^{-1}
\end{equation}
(\cite{dg84}) where $M_B$ and $\dot M_B$ are the total mass 
and mass loss rate of the blob, $\kappa_{class}$ is the 
classical thermal conductivity (\cite{spitzer62}), $R_B 
= 10^7 \, R_7$~cm is the blob radius, and $n_B = 10^{20} 
\, n_{20}$~cm$^{-3}$ is the particle density in the blobs. 
Eq. (1) indicates that for $n_{20} \, R_7^2 \gtrsim 0.1$
the blobs are expected to drift through the corona without 
significant mass-loss.

\section{Summary and conclusions}

We propose a new model for the rapid aperiodic variability of
the X-ray emission of GBHCs, involving a cool, dense blob drifting 
inward through a hot, inhomogeneous Comptonizing corona and disappearing 
at the event horizon of a solar-mass black hole. We have demonstrated 
that the resulting spectral and temporal characteristics predicted by 
this model are in very good agreement with the observed X-ray 
spectra, power spectra, hard time lags and coherence functions
of Cyg~X-1 and GX~339-4 if the hot corona has a radial temperature 
and density profile corresponding to an ADAF and the blob heats 
up slowly ($T_{BB} \propto r^{-1/4}$) as it moves inward. 

We have specified the conditions under which such cool blobs
are stable against evaporation on the radial drift time scale
of $\sim 1$~s, which is the case for $n_{20} \, R_7^2 \gtrsim 
0.1$. From a comparison with observed power spectra of GBHCs we 
can deduce that the power spectrum of blob formation events
must be given by a $\nu_F^{-1}$ power law, and that these
events occur on a typical repetition time scale $\sim 
1$~s~$\lesssim \Delta t_{rep} \lesssim 10$~s.

Our model is in accord with the increasing evidence, from 
observational as well as theoretical results, that accretion
onto Galactic black holes does not occur in a quasi-static
equilibrium state, but that accretion instabilities lead to
a rather erratic, inhomogeneous structure of inner accretion 
flows. The inward-motion of localized, efficiently radiating 
cool blobs through an inhomogeneous, hot corona then provides 
a very natural explanation for the observed hard time lags 
in the aperiodic variability of GBHCs.

\acknowledgements{We thank the anonymous referee vor very
helpful and constructive comments. This work is partially 
supported by NASA grant NAG~5-4055.}

\newpage

\begin{figure}
\rotate[r]{
\epsfysize=12cm
\epsffile[150 0 550 500]{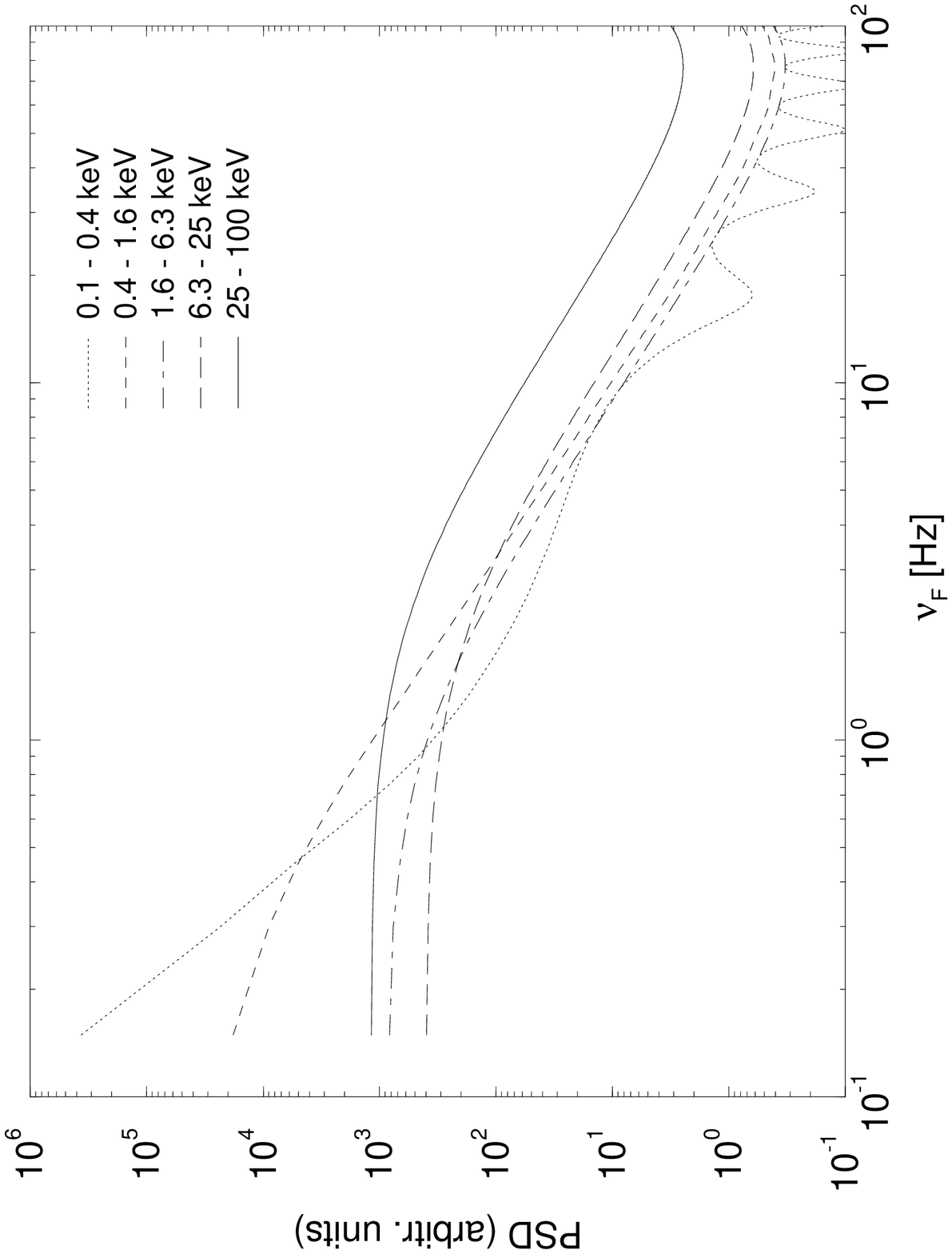}
}
\caption[]{Power spectra of light curves in five photon energy 
bins for a blob drifting inward with $\beta_r = 5 \cdot 10^{-3}$ 
through an ADAF-like corona and heating up according to $T_{BB} 
\propto r^{-1/4}$, $kT_{BB} (r_{in}) = 0.2$~keV. Coronal parameters: 
$k T_c (r_{in}) = 300$~keV, $T_c \propto r^{-1}$, $n_c \propto 
r^{-3/2}$, $\tau_c = 3$, $r_{in} = 10^7$~cm, $r_{out} = 10^9$~cm.}
\end{figure}

\newpage

\begin{figure}
\rotate[r]{
\epsfysize=12cm
\epsffile[150 0 550 500]{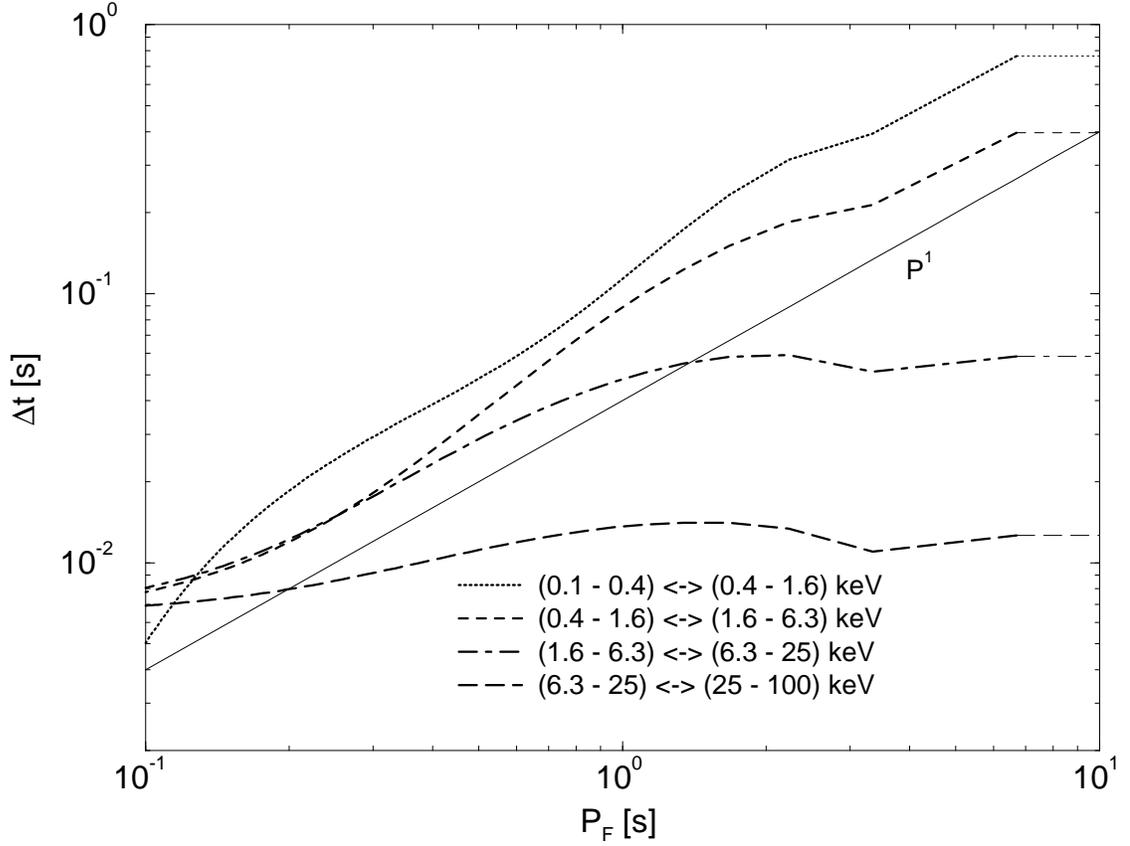}
}
\caption[]{Hard time lags between adjacent photon energy bins
as a function of Fourier period $P_F$; for parameters see caption 
of Fig. 1. Thin curves are the theoretical continuation of the
phase lag curves at long Fourier periods which are not covered
by our simulations. }
\end{figure}

\end{document}